\documentclass[fleqn]{article}
\usepackage{amsmath,amssymb}
\usepackage{epsfig,graphicx}

\catcode`@=11 \@addtoreset{equation}{section} \catcode`@=12

\begin{document}
\title{\vskip-1.7cm \bf  Anomaly Driven Cosmology: Big Boost Scenario and AdS/CFT
Correspondence}
\date{}
\author{A.O.Barvinsky$^{1}$, C.Deffayet$^2$ and A.Yu.Kamenshchik$^{3,\,4}$}
\maketitle \hspace{-8mm} {\,\,$^{1}$\em Theory Department, Lebedev
Physics Institute, Leninsky Prospect 53, Moscow 119991, Russia\\
$^{2}$APC UMR 7164 (CNRS, Univ. Paris 7, CEA, Obs. de Paris), 12 rue
Alice Domon et L\'eonie Duquet, 75205 Paris Cedex 13, France\\
$^{3}$Dipartimento di Fisica and INFN, via
Irnerio 46, 40126
Bologna, Italy\\
$^{4}$L.D.Landau Institute for Theoretical Physcis of Russian
Academy of Sciences, Kosygin str. 2, 119334 Moscow, Russia}

\begin{abstract}
We consider the cosmological evolution in a recently suggested new
model of quantum initial conditions for the Universe. The effective
Friedmann equation incorporates the effect of the conformal anomaly
of quantum fields and, interestingly, shows that their vacuum
Casimir energy is completely screened and does not gravitate. The
cosmological evolution also features  a new mechanism for a
cosmological acceleration stage. This stage is followed by a {\it
big boost} singularity --- a rapid growth up to infinity of the
scale factor acceleration parameter. We also briefly discuss the
relation between our model, the AdS/CFT correspondence and RS and
DGP braneworlds.
\end{abstract}

\section{Introduction}

Conformally invariant fields play a very important role in physics.
In particular, the success of string theory is largely due to the
fact that local conformal invariance provides us with the means of
exactly solving the underlying 2D problem. Conformal invariance is
pertinent not only to deal with the 2D dynamics on the string world
sheet, but also manifests itself in field theoretical implications
of string theory like the AdS/CFT correspondence suggesting new
calculational methods to describe a strongly coupled regime. On the
other hand, conformally invariant fields have important implications
in cosmology of the early Universe. Cosmological evolution driven by
the conformal anomaly of conformally invariant fields
\cite{FHH,Starobinsky} represented perhaps first examples of
self-consistent inflationary models. These earlier works disregarded
the formulation of initial conditions which later were considered in
the form of the no-boundary proposal \cite{HH} within the conformal
anomaly context and, more recently, using the AdS/CFT correspondence
\cite{HHR}.

Some of these ideas were recently generalized in a new model of
quantum initial conditions for the Universe \cite{slih}. In this
model, one considers a (possibly large) number of conformal fields
as the initial matter content of the Universe, also endowed with a
cosmological constant, and allows the possibility that the initial
state of the Universe be a mixed state, rather that a pure state as
in the Harle-Hawking proposal. Using the conformal invariance, it is
then possible in this model to compute the statistical sum in
quantum gravity of spatially closed cosmologies \cite{slih}. Indeed,
the initial state is represented by the microcanonical density
matrix \cite{why} whose statistical sum can be calculated within the
$1/N_{\rm cdf}$-expansion in the number, $N_{\rm cdf}$, of
conformally invariant quantum fields under the assumption that they
outnumber all other degrees of freedom. This statistical sum is
dominated by the set of the quasi-thermal cosmological instantons. A
first very interesting outcome of this calculation is the fact that
its consistency requires the effective cosmological constant of the
early Universe to belong to a finite range, strictly bounded from
above and from below \cite{slih}. It also shows that the vacuum
Hartle-Hawking instantons are excluded from the initial conditions,
having {\em infinite positive} Euclidean gravitational effective
action \cite{slih} due to the contribution of the conformal anomaly.

Here we are going to analyze the cosmological evolution in this
model with the initial conditions set by the instantons of
\cite{slih}. In particular, we will derive the modified Friedmann
equation incorporating the effect of conformal anomaly at late
radiation and matter domination stages. As will be shown, this
equation has several interesting properties. First it shows that the
vacuum (Casimir) part of the energy density is "degravitated" via
the effect of the conformal anomaly. Namely the Casimir energy does
not weigh. Second, we will show that this equation, together with
the recovery of the general relativistic behavior, can feature a
stage of cosmological acceleration followed by what we call a {\em
big boost} singularity. At this singularity the scale factor
acceleration grows in finite proper time up to infinity with a
finite limiting value of the Hubble factor, when the Universe again
enters a quantum phase demanding for its description {\bf a} UV
completion of the low-energy semiclassical theory. Finally we
discuss the possibility of realizing this scenario within the
AdS/CFT and braneworld setups, in particular when the conformal
anomaly and its effective action are induced on 4D boundary/brane
from the type IIB supergravity theory in the 5D bulk. We also
comment on the relation between our model and the DGP setup.

\section{Cosmological initial conditions generated by the conformal
anomaly}

The statistical sum for the microcanonical ensemble in spatially
closed cosmology ($S^3$-topology of spatial sections) was shown
to be represented by the path integral over the periodic scale
factor $a(\tau)$ and lapse function $N(\tau)$ of the minisuperspace
metric
    \begin{eqnarray}
    ds^2 = N^2(\tau)\,d\tau^2+a^2(\tau)\,d^2\Omega^{(3)} \label{FRW}
    \end{eqnarray}
on the toroidal spacetime of $S^1\times S^3$ topology \cite{why}
    \begin{eqnarray}
    e^{-\varGamma}=\!\!\int\limits_{\,\,\rm periodic}
    \!\!\!\! D[\,a,N\,]\;
    e^{-\varGamma_E[\,a,\,N\,]}.   \label{1}
    \end{eqnarray}
Here $\varGamma_E[\,a,\,N]$ is the Euclidean effective action of all
inhomogeneous ``matter" fields which include also metric
perturbations on minisuperspace background of (\ref{FRW}).

Under the assumption that the system is dominated by free matter
fields conformally coupled to gravity this action is exactly
calculable by the conformal transform from (\ref{FRW}) to static
Einstein metric with $a={\rm const}$ \cite{slih}. In units of the
Planck mass $m_P=(3\pi/4G)^{1/2}$ it reads
    \begin{eqnarray}
    &&\varGamma_E[\,a,N\,]=m_P^2\int d\tau\,N \left\{-aa'^2
    -a+ \frac\Lambda3 a^3
    +B\!\left(\frac{a'^2}{a}              \label{FrieEu}
    -\frac{a'^4}{6 a}\right)
    +\frac{B}{2a}\,\right\}+ F(\eta),\\
    &&F(\eta)=\pm\sum_{\omega}\ln\big(1\mp
    e^{-\omega\eta}\big),\,\,\,\,\,
    \eta=\int d\tau N/a.                 \label{effaction}
    \end{eqnarray}
Here $a'\equiv da/Nd\tau$, the first three terms in curly brackets
represent the classical Einstein action with a primordial
cosmological constant $\Lambda$, the $B$-terms correspond to the
contribution of the conformal anomaly and the contribution of the
vacuum (Casimir) energy $(B/2a)$ of conformal fields on a static
Einstein spacetime. $F(\eta)$ is a free energy of these fields -- a
typical boson or fermion sum over field oscillators with energies
$\omega$ on a unit 3-sphere, $\eta$ playing the role of the inverse
temperature --- an overall circumference of the toroidal instanton
measured in units of the conformal time. The constant $B$,
    \begin{eqnarray}
    B=\frac{3\beta}{4 m_P^2}=\frac{\beta G}\pi,   \label{B}
    \end{eqnarray}
is determined by the coefficient $\beta$ of the topological
Gauss-Bonnet invariant $E = R_{\mu\nu\alpha\gamma}^2-4R_{\mu\nu}^2 +
R^2$ in the overall conformal anomaly of quantum fields
    \begin{equation}
    g_{\mu\nu}\frac{\delta
    \varGamma_E}{\delta g_{\mu\nu}} =
    \frac{1}{4(4\pi)^2}g^{1/2}
    \left(\alpha \Box R +
    \beta E + \gamma C_{\mu\nu\alpha\beta}^2\right),    \label{anomaly}
    \end{equation}
($C^2_{\mu\nu\alpha\beta}$ is the Weyl tensor squared term). For the
model of $N_0$ scalar, $N_{1/2}$ Weyl spinor and $N_{1}$ gauge
vector fields it reads \cite{Duffanomaly}
    \begin{eqnarray}
    \beta=\frac1{360}\,\big(2 N_0+11 N_{1/2}+
    124 N_{1}\big).                \label{100}
    \end{eqnarray}

The coefficient $\gamma$ does not contribute to (\ref{FrieEu})
because the Weyl tensor vanishes for any FRW metric. The situation
with the coefficient $\alpha$ is more complicated. A nonvanishing
$\alpha$ induces higher derivative terms $\sim \alpha (a'')^2$ in
the action and, therefore, adds one extra degree of freedom to the
minisuperspace sector of $a$ and $N$ and results in
instabilities\footnote{In Einstein theory this sector does not
contain physical degrees of freedom at all, which solves the problem
of the formal ghost nature of $a$ in the Einstein Lagrangian.
Addition of higher derivative term for $a$ does not formally lead to
a ghost -- the additional degree of freedom has a good sign of the
kinetic term as it happens in $f(R)$-theories, but still leads to
instabilities discovered in \cite{Starobinsky}.}. But $\alpha$ can
be renormalized to zero by adding a finite {\em local} counterterm
$\sim R^2$ admissible by the renormalization theory. We assume this
{\em number of degrees of freedom preserving} renormalization to
keep theory consistent both at the classical and quantum levels
\cite{slih}. It is interesting that this finite renormalization
changes the value of the Casimir energy of conformal fields in
closed Einstein cosmology in such a way that for all spins this
energy is universally expressed in terms of the same conformal
anomaly coefficient $B$ (corresponding to the $B/2a$ term in
(\ref{FrieEu})) \cite{slih}. As we will see, this leads to the
gravitational screening of the Casimir energy, mediated by the
conformal anomaly of quantum fields.

Ultimately, the effective action (\ref{FrieEu}) contains only two
dimensional constants -- the Planck mass squared (or the
gravitational constant) $m_P^2=3\pi/4G$ and the cosmological
constant $\Lambda$. They have to be considered as renormalized
quantities. Indeed, the effective action of conformal fields
contains divergences, the quartic and quadratic ones being absorbed
by the renormalization of the initially singular bare cosmological
and gravitational constants to yield finite renormalized $m_P^2$ and
$\Lambda$. Logarithmically divergent counterterms have the same
structure as curvature invariants in the anomaly (\ref{anomaly}).
When integrated over the spacetime closed toroidal FRW instantons
they identically vanish because the $\Box R$ is a total derivative,
Euler number of $S^3\times S^1$ is zero, $\int d^4x g^{1/2}E=0$, and
$C_{\mu\nu\alpha\beta}=0$. There is however a finite tail of these
vanishing logarithmic divergences in the form of the conformal
anomaly action which incorporates the coefficient $\beta$ of $E$ in
(\ref{anomaly}) and constitutes a major contribution to
$\varGamma_E$ --- the first two $B$-dependent terms of
(\ref{effaction})\footnote{These terms can be derived from the
metric-dependent Riegert action \cite{Riegert} or the action in
terms of the conformal factor relating two metrics \cite{FrTs,BMZ}
and generalize the action of \cite{FHH} to the case of a spatially
closed cosmology with $\alpha=0$.}. Thus, in fact, this model when
considered in the leading order of the $1/N$-expansion (therefore
disregarding loop effects of the graviton and other non-conformal
fields) is renormalizable in the minisuperspace sector of the
theory.

The path integral (\ref{1}) is dominated by the saddle points ---
solutions of the equation $\delta\varGamma_E/\delta N(\tau)=0$ which
reads as
    \begin{eqnarray}
    &&-\frac{a'^2}{a^2}+\frac{1}{a^2}
    -B \left(\frac12\,\frac{a'^4}{a^4}
    -\frac{a'^2}{a^4}\right) =
    \frac\Lambda3+\frac{C}{ a^4},     \label{efeq}\\
    &&C = \frac{B}2 +\frac{dF(\eta)}{d\eta},\,\,\,\,
    \eta = 2k \int_{\tau_-}^{\tau_+}
    \frac{d\tau}{a}.                       \label{bootstrap}
    \end{eqnarray}
Note that the usual (Euclidean) Friedmann equation is modified by
the anomalous $B$-term and the radiation term $C/a^4$. The constant
$C$ sets the amount of radiation and satisfies the bootstrap
equation (\ref{bootstrap}), where $B/2$ is a contribution of the
Casimir energy, and
    \begin{eqnarray}
    \frac{dF(\eta)}{d\eta}=
    \sum_\omega\frac{\omega}{e^{\omega\eta}\mp 1}        \label{100}
    \end{eqnarray}
is the energy of the gas of thermally excited particles with the
inverse temperature $\eta$. The latter is given in (\ref{bootstrap})
by the $k$-fold integral between the turning points of the scale
factor history $a(\tau)$, $\dot a(\tau_\pm)=0$. This $k$-fold nature
implies that in the periodic solution the scale factor $k$ times
oscillates between its maximum and minimum values
$a_\pm=a(\tau_\pm)$.

As shown in \cite{slih}, such solutions represent the garland-type
instantons which exist only in the limited range of the cosmological
constant
    \begin{eqnarray}
    0<\Lambda_{\rm min}<\Lambda<
    \frac{3\pi}{2\beta G},                \label{landscape}
    \end{eqnarray}
and eliminate the vacuum Hartle-Hawking instantons corresponding to
$a_-=0$\footnote{Hartle-Hawking instantons are ruled out in the
statistical sum by their infinite positive effective action which is
due to the contribution of the conformal anomaly drastically
changing predictions of the tree-level theory.}. The period of these
quasi-thermal instantons is not a freely specifiable parameter, but
as a function of $G$ and $\Lambda$ follows from Eqs.
(\ref{efeq})-(\ref{bootstrap}). Therefore the model does not
describe a canonical ensemble, but rather a microcanonical ensemble
(see \cite{why}) with only two freely specifiable dimensional
parameters  --- the renormalized gravitational and cosmological
constants discussed above.

The upper bound of the range (\ref{landscape}) is entirely caused by
the quantum anomaly -- this is a new quantum gravity scale which
tends to infinity when one switches the quantum effects off,
$\beta\to 0$. The lower bound $\Lambda_{\rm min}$ is the effect of
both radiation and anomaly, and can be obtained numerically for any
field contents of the model. For a large number of conformal fields,
and therefore a large $\beta$, the lower bound is of the order
$\Lambda_{\rm min}\sim 1/\beta G$. Thus the restriction
(\ref{landscape}) can be regarded as a solution of the cosmological
constant problem in the early Universe, because specifying a
sufficiently high number of conformal fields one can achieve the
primordial value of $\Lambda$ well below the Planck scale where the
effective theory applies, but high enough to generate sufficiently
long inflationary stage. Also this restriction can be potentially
considered as a selection criterion for the landscape of string
vacua \cite{slih,why}.

\section{Cosmological evolution}

\begin{figure}[h]
\centerline{\epsfxsize 4.4cm \epsfbox{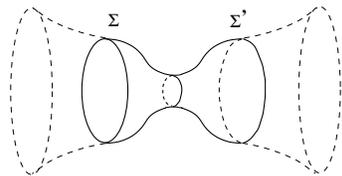}} \caption{\small The
contribution of the one-fold instanton to the density matrix of the
Universe, whose two arguments are associated with with the surfaces
$\Sigma$ and $\Sigma'$. Dashed lines depict the Lorentzian Universe
nucleating at $\Sigma$ and $\Sigma'$. \label{Fig.1}}
\end{figure}

The gravitational instantons of the above type can be regarded as
setting initial conditions for the cosmological evolution in the
physical spacetime with the Lorentzian signature. This can be viewed
as nucleation of the Lorentzian spacetime from the Euclidean
spacetime at the maximum value of the scale factor $a_+=a(\tau_+)$
at the turning point of the Euclidean solution $\tau_+$ --- the
minimal (zero extrinsic curvature) surface of the instanton. For the
contribution of the one-fold instanton to the density matrix of the
Universe this nucleation process is depicted in Fig. \ref{Fig.1}.

The Lorentzian evolution can be obtained by analytically continuing
the Euclidean time into a complex plane by the rule
$\tau=\tau_++it$. Correspondingly the Lorentzian effective equation
follows from the Euclidean one (\ref{efeq}) as
    \begin{eqnarray}
    &&\frac{\dot a^2}{a^2}+\frac{1}{a^2}
    -B \left(\frac12\,\frac{\dot a^4}{a^4}
    +\frac{\dot a^2}{a^4}\right) =
    \frac\Lambda3+\frac{C}{ a^4},
    \end{eqnarray}
where the dot from now on denotes the derivative with respect to the
Lorentzian time $t$. This can be rewritten in the form
    \begin{eqnarray}
    \frac{\dot a^2}{a^2}+\frac{1}{a^2}-
    \frac{B}2 \left(\frac{\dot a^2}{a^2}
    +\frac{1}{a^2}\right)^2 =
    \frac\Lambda3+\frac{C-B/2}{ a^4},   \label{Friedmann3}
    \end{eqnarray}
and solved for the Hubble factor as
    \begin{eqnarray}
    &&\frac{\dot
    a^2}{a^2}+\frac{1}{a^2}=
    \frac1B\left\{1-
    \sqrt{1-2B\left(\frac\Lambda3
    +\frac{\cal C}{a^4}\right)}\right\},         \label{Friedmann0}\\
    &&{\cal C} \equiv C-\frac{B}2.              \label{calC}
    \end{eqnarray}
We have thus obtained a modified Friedmann equation in which the
overall energy density, including both the cosmological constant and
radiation, very nonlinearly contributes to the square of the Hubble
factor.

An interesting property of this equation is that the Casimir energy
does not weigh --- the term $B/2a^4$ is completely subtracted
from the full radiation density $C/a^4$ in the right hand side of
(\ref{Friedmann3}) and under the square root of (\ref{Friedmann0}).
Only ``real" thermally excited quanta contribute to the right-hand
side of (\ref{Friedmann0}). Indeed, using (\ref{bootstrap}), the
radiation contribution ${\cal C}/a^4$ is seen to read simply
as
    \begin{eqnarray}
    \frac{\cal C}{a^4} = \frac1{a^4}
    \sum_\omega\frac{\omega}{e^{\omega\eta}\mp 1}. \label{primrad}
    \end{eqnarray}
This is an example of the gravitational screening which is now being
intensively searched for the cosmological constant
\cite{WoodardTsamis,DvaliKhouryetal}. As we see this mechanism is
mediated by the conformal anomaly action, but it applies not to the
cosmological constant, but rather to the Casimir energy which has
the equation of state of radiation $p=\varepsilon/3$. This
gravitational screening is essentially based on the above mentioned
renormalization that eradicates higher derivatives from the
effective action and thus preserves the minisuperspace sector free
from dynamical degrees of freedom.

After nucleation from the Euclidean instanton at the turning point
with $a=a_+$ and $\dot a_+=0$ the Lorentzian Universe starts
expanding, because
    \begin{eqnarray}
    \ddot a_+=-a\left.\frac{1+
    \dot a^2-2\Lambda a^2/3}{a^2-B-B\dot
    a^2}\right|_{\,\tau_+}
    =a_+\frac{\sqrt{1-4C\Lambda/3}}{a_+^2-B}>0   \label{nucleation}
    \end{eqnarray}
(this equation can be derived from (\ref{Friedmann0}) or obtained by
analytic continuation from the Euclidean variational equation
$\delta \varGamma_E[a,N]/\delta a=0$). Therefore, the radiation
quickly dilutes, so that the primordial cosmological constant starts
dominating and can generate an inflationary stage. It is natural to
assume that the primordial $\Lambda$ is not fundamental, but is due
to some inflaton field. This effective $\Lambda$ is nearly constant
during the Euclidean stage and the inflation stage, and subsequently
leads to a conventional exit from inflation by the slow roll
mechanism\footnote{In the Euclidean regime this field also stays in
the slow roll approximation, but in view of the oscillating nature
of a scale factor it does not monotonically decay. Rather it follows
these oscillations with much lower amplitude and remains nearly
constant during all Euclidean evolution, whatever long this
evolution is (as it happens for garland instantons with the number
of folds $k\to\infty$).}.

During a sufficiently long inflationary stage, particle production
of conformally non-invariant matter takes over the polarization
effects of conformal fields. After being thermalized at the exit
from inflation this matter gives rise to an energy density
$\varepsilon(a)$ which should replace the energy density of the
primordial cosmological constant and radiation. Therefore, at the
end of inflation the combination $\Lambda/3+{\cal C}/a^4$ should be
replaced according to
    \begin{eqnarray}
    \frac\Lambda3+\frac{\cal C}{a^4}\to
    \frac{8\pi G}3\,\varepsilon(a)\equiv
    \frac{8\pi G}3\,\rho(a)+\frac{\cal C}{a^4}.   \label{split}
    \end{eqnarray}
Here $\varepsilon(a)$ denotes the full energy density including
the component $\rho(a)$ resulting from the decay of $\Lambda$ and
the radiation density of the primordial conformal matter ${\cal
C}/a^4$. The dependence of $\varepsilon(a)$ on $a$ is of course
determined by the equation of state via the stress tensor
conservation, and $\rho(a)$ also includes its own radiation
component emitted by and staying in (quasi)equilibrium with the
barionic part of the full $\varepsilon(a)$.

Thus the modified Friedman equation finally takes the form
    \begin{eqnarray}
    \frac{\dot a^2}{a^2}+\frac{1}{a^2}=
    \frac\pi{\beta G}\left\{\,1-
    \sqrt{\,1-\frac{16 G^2}3\,
    \beta\varepsilon}\,\right\},              \label{modFriedmann}
    \end{eqnarray}
where we expressed $B$ according to (\ref{B})

In the limit of small subplanckian energy density $\beta
G^2\varepsilon\equiv\beta\varepsilon/\varepsilon_P\ll 1$ the
modified equation goes over into the ordinary Friedmann equation in
which the parameter $\beta$ completely drops out
    \begin{eqnarray}
    \frac{\dot a^2}{a^2}+\frac{1}{a^2}
    =\frac{8\pi
    G}3\,\varepsilon,\,\,\,\,
    G^2\varepsilon\ll\frac1\beta.     \label{GR}
    \end{eqnarray}
Therefore within this energy range the standard cosmology is
recovered. Depending on the effective equation of state, a wide set
of the known scenarios of late cosmological evolution can be
obtained, including quintessence \cite{quintessence} and other
scenarios of cosmological acceleration \cite{otherDE}.

\section{Big boost cosmological acceleration}

The range of applicability of the GR limit (\ref{GR}) depends
however on $\beta$. This makes possible a very interesting mechanism
to happen for a very large $\beta$. Indeed, the value of the
argument of the square root in (\ref{modFriedmann}) can be
sufficiently far from 1 even for small $\varepsilon$ provided
$\beta\sim N_{\rm cdf}\gg 1$. Moreover, one can imagine a model with
a variable number of the conformal fields $N_{\rm cdf}(t)$ inducing
a time and implicitly a scale factor dependent $\beta$,
$\beta=\beta(a)$. If $\beta(a)$ grows with $a$ faster than the rate
of decrease of $\varepsilon(a)$, then the solution of
(\ref{modFriedmann}) can reach the singular point labeled below by
$\infty$ at which the square root argument vanishes and the
cosmological acceleration becomes infinite. This follows from the
expression
    \begin{eqnarray}
    \frac{\ddot a}a=\frac\pi{\beta G}
    \left\{\left(1-\sqrt{\,1-16 G^2
    \beta\varepsilon/3}\,\right)
    \left(1-\frac12\frac{a(G\beta)'}{G\beta}\right)+
    \frac43\frac{a(G^2\beta\varepsilon)'}{\sqrt{\,1-16 G^2
    \beta\varepsilon/3}}\right\},     \label{acceleration}
    \end{eqnarray}
where prime denotes the derivative with respect to $a$. This
expression becomes singular at $t=t_\infty$ even though the Hubble
factor remains finite when
    \begin{eqnarray}
    &&(G^2\beta\varepsilon)_\infty=\frac3{16},\\
    &&\left(\frac{\dot a^2}{a^2}+\frac{1}{a^2}\right)_\infty
    =\frac{16\pi}3
    (G\varepsilon)_\infty.   \label{boostHubble}
    \end{eqnarray}

Assuming for simplicity that the matter density has a dust-like
behavior and $\beta$ grows by a power law in $a$
    \begin{eqnarray}
    G\varepsilon\sim \frac1{a^3},\,\,\,\,
    G\beta\sim
    a^n,\,\,\,\,n>3,              \label{behavior}
    \end{eqnarray}
one easily finds an inflection point $t=t_*$ when the cosmological
acceleration starts after the deceleration stage
    \begin{eqnarray}
    &&(G^2\beta\varepsilon)_*=\frac3{4}\frac{n-2}{(n-1)^2},\\
    &&\left(\frac{\dot a^2}{a^2}+\frac{1}{a^2}\right)_*
    =\frac{8\pi}3
    (G\varepsilon)_*\frac{n-1}{n-2}.    \label{H*}
    \end{eqnarray}
Of course, for the acceleration stage to start, the Universe should
reach this inflection point $t_*$ before recollapsing from the point
of its maximal expansion. This requirement imposes certain
restrictions on the coefficients of the asymptotic behavior of
$G\varepsilon$ and $G\beta$, (\ref{behavior}), which depend on the
details of the history of the equation of state for $\varepsilon(a)$
and the dynamics in the number of conformal fields. We will consider
these details elsewhere.

 Here it is worth only mentioning that the matter density in
the vicinity of $t_*$ gravitates with {\bf a} slightly rescaled
gravitational constant, Eq.(\ref{H*}), while at the singularity the
effective gravitational constant doubles, see
Eq.(\ref{boostHubble}). Also it is useful to comment on the
duration of the acceleration stage before reaching the singularity.
If we identify our epoch with some instant $t_0$ soon after $t_*$,
then this duration until the singularity can be estimated by
disregarding the spatial curvature term. Then it reads as
    \begin{eqnarray}
    t_\infty-t_*\sim\sqrt{B_0}\sim H_0^{-1},
    \end{eqnarray}
which is comparable to the age of the Universe. Thus, although the
acceleration stage does not pass the eternity test of
\cite{Polyakov}, its duration is very large.

Nevertheless, the evolution ends in this model with the curvature
singularity, $\ddot a\to\infty$, reachable in a finite proper time.
Unlike the situation with a big brake singularity of \cite{bigbrake}
it cannot be extended beyond this singularity analytically even by
smearing it or taking into account its weak integrable nature. In
contrast to \cite{bigbrake} the acceleration is positive, which fact
allows us to call this singularity a {\em big boost}. The effect of
the conformal anomaly drives the expansion of the Universe to the
maximum value of the Hubble constant (\ref{boostHubble}), after
which the solution becomes complex. This, of course, does not make
the model apriori inconsistent, because for $t\to t_*$ an infinitely
growing curvature invalidates the semiclassical and $1/N$
approximations. This is a new essentially quantum stage which
requires the UV completion of the effective low-energy theory.

\section{AdS/CFT and Randall-Sundrum braneworld setup}

What can be the mechanism of a variable and indefinitely growing
$\beta$? One such mechanism is well known -- phase transitions in
cosmology between different vacua can give a mass $m$ to an
initially massless particle. This results in the loss of conformal
invariance of the corresponding particle, which instead of
contributing to the vacuum polarization by its own $\beta$ factor
starts generating the Coleman-Weinberg type potential $\sim
m^4\ln(m^2/\mu^2)$. However this effect is weak and decreases the
effective value of $\beta$, which is not what we are after.

Another mechanism was suggested in \cite{why}. It relies on the
possible existence, motivated by string theory, of extra dimensions
whose size is evolving in time. Indeed, theories with extra
dimensions provide a qualitative mechanism to promote $\beta$ to the
level of a moduli variable indefinitely growing with the evolving
size $L$ of those dimensions. Indeed $\beta$ basically counts
the number $N_{\rm cdf}$ of conformal degrees of freedom, $\beta\sim
N_{\rm cdf}$ (see Eq.(\ref{100})). If one considers a string theory
in a space time with more than four dimensions, the extra-dimension
being compact with typical size $L$, the effective 4-dimensional
fields arise as Kaluza-Klein (KK) and winding modes with masses (see
e.g. \cite{Polch})
    \begin{eqnarray}
    m_{n,w}^2=\frac{n^2}{L^2}+\frac{w^2}{\alpha'^2}\,L^2
    \end{eqnarray}
(where $n$ and $w$ are respectively the KK and winding numbers),
which break their conformal symmetry. These modes remain
approximately conformally invariant as long as their masses are much
smaller than the spacetime curvature, $m_{n,w}^2\ll H_0^2\sim
m_P^2/N_{\rm cdf}$. Therefore the number of conformally invariant
modes changes with $L$. Simple estimates show that the number of
pure KK modes ($w=0$, $n\leq N_{\rm cdf}$) grows with $L$ as $N_{\rm
cdf}\sim (m_P L)^{2/3}$, whereas the number of pure winding modes
($n=0$, $w\leq N_{\rm cdf}$) grows as $L$ decreases as $N_{\rm
cdf}\sim(m_P\alpha'/L)^{2/3}$. Thus, it is possible to find a
growing $\beta$ in both cases with expanding or contracting extra
dimensions. In the first case it is the growing tower of
superhorizon KK modes which {\it makes} the horizon scale $H_0\sim
m_P/\sqrt{N_{\rm cdf}}\sim m_P/(m_P L)^{1/3}$ indefinitely decrease
with $L\to\infty$. In the second case it is the tower of
superhorizon winding modes which makes this scale decrease with the
decreasing $L$ as $H_0\sim m_P(L/m_P\alpha')^{1/3}$. At the
qualitative level of this discussion so far, such a scenario is
flexible enough to accommodate the present day acceleration scale
(though, at the price of fine-tuning an enormous coefficient of
expansion or contraction of $L$).

However, string (or rather string-inspired) models can offer a more
explicit construction of the ideas put forward previously, as well
as help addressing various phenomenological questions which arise in
their consideration. In particular, one obvious question which
arises, should the model considered here be realistic, is what are
the possible observable effects of the large number of required
conformal fields. Here some guidance can be obtained from the
AdS/CFT picture. Indeed, in this picture \cite{AdS/CFT}
a higher dimensional theory of gravity, namely type IIB supergravity
compactified on $AdS_5 \times S^5$, is seen to be equivalent to a
four dimensional conformal theory, namely  ${\cal N}=4$ $SU(N)$ SYM,
thought to live on the boundary of $AdS_5$ space-time. An
interesting arena for a slight generalization of these ideas is the
Randall-Sundrum model \cite{Randall:1999vf} where a 3-brane is put
in the inside of $AdS_5$ space-time resulting in a large distance
recovery of 4D gravity without the need for compactification. This
model has a dual description. On the one hand it can just be
considered from a 5D gravity perspective, on the other hand it can
also be described, thanks to the AdS/CFT picture, by a 4D conformal
field theory coupled to gravity.

Indeed, in this picture, the 5D SUGRA --- a field-theoretic limit of
the type IIB string --- induces on the conformal boundary of the
underlying AdS background the quantum effective action of the
conformally invariant 4D ${\cal N}=4$ $SU(N)$ SYM theory coupled to
the 4D geometry of the boundary. The multiplets of this CFT
contributing according to (\ref{100}) to the total conformal anomaly
coefficient $\beta$ are given by $(N_0,N_{1/2},N_1)=(6N^2,4N^2,N^2)$
\cite{DuffLiu}, so that
    \begin{eqnarray}
    \beta=\frac12\,N^2.
    \end{eqnarray}
The parameters of the two theories are related by the equation
\cite{AdS/CFT,Gubser,HHR}
    \begin{eqnarray}
    \frac{L^3}{2 G_5}=\frac{N^2}{\pi},
    \end{eqnarray}
where $L$ is the radius of the 5D $AdS$ space-time with the negative
cosmological constant $\Lambda_5=-6/L^2$ and $G_5$ is the 5D
gravitational constant. The radius $L$ is also related to the 't
Hooft parameter of the SYM coupling $\lambda=g_{SYM}^2 N$ and the
string length scale $l_s=\sqrt{\alpha'}$, $L=\lambda^{1/4} l_s$. The
generation of the 4D CFT from the local 5D supergravity holds in the
limit when both $N$ and $\lambda$ are large. This guarantees the
smallness of string corrections and establishes the relation between
the weakly coupled tree-level gravity theory in the bulk ($G_5\to
0$, $L\to\infty$) and the strongly coupled 4D CFT ($g_{SYM}^2\gg
1$). Moreover, as said above, the AdS/CFT correspondence explains
the mechanism of recovering general relativity theory on the 4D
brane of the Randall-Sundrum model \cite{Gubser,HHR}. The 4D gravity
theory is induced on the brane from the 5D theory with the negative
cosmological constant $\Lambda_5=-6/L^2$. In the one-sided version
of this model the brane has a tension $\sigma=3/8\pi G_5L$ (the 4D
cosmological constant is given by $\Lambda_4=8\pi G_4\sigma$), and
the 4D gravitational constant $G_4\equiv G$ turns out to be
    \begin{eqnarray}
    G=\frac{2G_5}L.
    \end{eqnarray}
One recovers 4D General Relativity at low energies and for distances
larger than the radius of the AdS bulk, $L$. Thus, the CFT dual
description of the 5D Randall-Sundrum model is very similar to the
model considered above. Moreover, even though the CFT effective
action is not exactly calculable for $g_{SYM}^2\gg 1$ it is
generally believed that its conformal anomaly is protected by
extended SUSY \cite{TseytlinLiu} and is exactly given by the
one-loop result (\ref{anomaly}). Therefore it generates the exact
effective action of the anomalous (conformal) degree of freedom
given by (\ref{effaction}), which guarantees a good $1/N_{\rm
cdf}$-approximation for the gravitational dynamics.

Applying further the above relations it follows a relation between
our $\beta$ coefficient and the radius $L$ of the $AdS$ space-time,
given by $\beta G=\pi L^2/2$. Introducing this in the modified
Friedmann equation (\ref{modFriedmann}), the latter becomes
explicitly depending on the size of the 5D AdS spacetime as given by
    \begin{eqnarray}
    \frac{\dot a^2}{a^2}+\frac{1}{a^2}=
    \frac2{L^2}\left\{\,1-
    \sqrt{\,1-L^2 \left(\frac{8\pi G}3\,
    \rho+\frac{\cal C}{a^4}\right)}\,\right\}, \label{modFriedmann2}
    \end{eqnarray}
where we have reintroduced the decomposition (\ref{split}) of the
full matter density into the decay product of the inflationary and
matter domination stages $\rho$ and the thermal excitations of the
primordial CFT (\ref{primrad}).

For low energy density, $GL^2\rho\ll 1$ and $L^2 {\cal C}/a^4\ll 1$,
in the approximation beyond the leading order, cf. Eq.(\ref{GR}),
this equation reads\footnote{We assume that the dark radiation
term is redshifted for growing $a$ faster than matter term and
expand to the second order in $\rho$, but the first order in ${\cal
C}$.}
    \begin{eqnarray}
    \frac{\dot a^2}{a^2}+\frac{1}{a^2}
    \simeq\frac{8\pi
    G}3\,\rho\,
    \left(1+\frac{2\pi GL^2}3\rho\right) + \frac{\cal C}{a^4}
    \end{eqnarray}
and coincides with the modified Friedmann equation in the
Randall-Sundrum model \cite{BinDefLan}
    \begin{eqnarray}
    \frac{\dot a^2}{a^2}+\frac{1}{a^2}
    =\frac{8\pi
    G}3\,\rho\,
    \left(1+\frac{\rho}{2\sigma}\right)+
    \frac{\cal C}{a^4},                      \label{RS}
    \end{eqnarray}
where $\sigma=3/8\pi G_5L=3/4\pi GL^2$ is the Randal-Sundrum brane
tension and ${\cal C}$ is the braneworld constant of motion
\cite{BinDefLan,bulkBH}. Note that the thermal radiation on the
brane (of non-Casimir energy nature) is equivalent to the mass of
the bulk black hole associated with this constant. This fact can be
regarded as another manifestation of the AdS/CFT correspondence in
view of the known duality between the bulk black hole and the
thermal CFT on the brane \cite{bulkBH}. Thus indeed anomaly driven
cosmology coincides with the Randall-Sundrum one in the limit of low
density of matter and radiation.

Interestingly, this comparison between our model and the
Randall-Sundrum framework also allows {\bf one} to have some insight
on the phenomenologically allowed physical scales. Indeed, it is
well known that the presence of an extra-dimension in the
Randall-Sundrum model, or in the dual language, that of the CFT,
manifests itself typically at distances lower that the $AdS$ radius
$L$. Hence, it is perfectly possible to have a large number of
conformal fields in the Universe, {\it \`a  la} Randall-Sundrum,
without noticing their presence in the everyday experiments,
provided $L$ is small enough. Moreover, if one uses the scenario of
\cite{slih} to set the initial conditions for inflation, it provides
an interesting connection between the Hubble radius of inflation,
given by eq. (\ref{landscape}), and the distance at which the
presence of the CFT would manifest itself in gravity experiments,
both being given by $L$. Last, it seems natural in a string theory
setting, to imagine that the $AdS$ radius $L$ can depend on time,
and hence on the scale factor.

In this case, assuming that the AdS/CFT picture still holds when $L$
is adiabatically evolving, one can consider the possibility that
$GL^2\varepsilon$ is large, and that $L^2(t)$ grows faster than
$G\varepsilon(t)$ decreases during the cosmological expansion. One
would then get the cosmological acceleration scenario of the above
type followed by the big boost singularity.

In this case, however, should this acceleration scenario correspond
to the present day accelerated expansion, $L$ should be of the order
of the present size of the Universe, i.e. $L^{-2}\sim H_0^2$. Since
the Randall-Sundrum mechanism recovers 4D GR only at distances
beyond the curvature radius of the AdS bulk, $r\gg L$, it means that
local gravitational physics of our model (\ref{modFriedmann2}) at
the acceleration stage is very different from the 4D general
relativity. Thus this mechanism can hardly be a good candidate for
generating dark energy in real cosmology.

\section{Anomaly driven cosmology and the DGP model}

It is interesting that there exists {\bf an} even more striking
example of a braneworld setup dual to our anomaly driven model. This
is the generalized DGP model \cite{DGP} including together with the
4D and 5D Einstein-Hilbert terms also the 5D cosmological constant,
$\Lambda_5$, in the special case of the {\em vacuum} state on the
brane with a vanishing matter density $\rho=0$. In contrast to the
Randall-Sundrum model, for which this duality holds only in the low
energy limit --- small $\rho$ and small ${\cal C}/a^4$, vacuum DGP
cosmology {\em exactly} corresponds to the model of \cite{slih} with
the 4D cosmological constant $\Lambda$ simulated by the 5D
cosmological constant $\Lambda_5$.

Indeed, in this model (provided one neglects the bulk
curvature), gravity interpolates between a 4D behaviour at small
distances and a 5D behaviour at large distances, with the crossover
scale between the two regimes being given by $r_c$,
    \begin{eqnarray}
    \frac{G_5}{2G}=r_c,      \label{DGPscale}
    \end{eqnarray}
and in the absence of stress-energy exchange between the brane and
the bulk, the modified Friedmann equation takes the form
\cite{DGPDeffayet}
    \begin{eqnarray}
    \frac{\dot a^2}{a^2}+\frac{1}{a^2}-
    r_c^2 \left(\,\frac{\dot a^2}{a^2}
    +\frac{1}{a^2}-\frac{8\pi G}3\,\rho\right)^2 =
    \frac{\Lambda_5}{6}
    +\frac{{\cal C}}{ a^4}.             \label{FriedmannDGP}
    \end{eqnarray}
Here ${\cal C}$ is the same as above constant of integration
of the bulk Einstein's equation, which corresponds to a
nonvanishing Weyl tensor in the bulk (or a mass for a Schwarzschild
geometry in the bulk) \cite{BinDefLan,bulkBH}. It is remarkable that
this equation with $\rho=0$ exactly coincides with the
modified Friedmann equation of the anomaly driven cosmology
(\ref{Friedmann3}) under the identifications
    \begin{eqnarray}
    &&B\equiv\frac{\beta G}\pi=2 r_c^2, \label{1000}\\
    &&\Lambda=\frac{\Lambda_5}2.
    \end{eqnarray}
These identifications imply that in the DGP limit $G\ll r_c^2$, the
anomaly coefficient $\beta$ is much larger than 1.

This looks very much like the generation of the vacuum DGP model for
any value of the dark radiation ${\cal C}/a^4$ from the anomaly
driven cosmology with a very large $\beta\sim m_P^2 r_c^2\gg 1$.
However, there are several differences. A first important difference
between the conventional DGP model and the anomaly driven DGP is
that the former does not incorporate the self-accelerating branch
\cite{DGPDeffayet,DDG} of the latter. This corresponds to the fact
that only one sign of the square root is admissible in
Eq.(\ref{Friedmann0}) --- a property dictated by the instanton
initial conditions at the nucleation of the Lorentzian spacetime
from the Euclidean one (see Eq.(\ref{nucleation})). So, one does not
have to worry about possible instabilities associated with the
self-accelerating branch \cite{instabilityofacceleratingbranch}.

 Another important difference concerns the way the matter energy
density manifests itself in the Friedmann equation for the
non-vacuum case.  In our 4D anomaly driven model it enters the right
hand side of the equation as a result of the decay (\ref{split}) of
the effective 4D cosmological constant $\Lambda$, while in the DGP
model it appears inside the parenthesis of the left hand side of
equation (\ref{FriedmannDGP}). Therefore, the DGP Hubble factor
reads as
    \begin{eqnarray}
    \frac{\dot a^2}{a^2}+\frac{1}{a^2}=
    \frac{8\pi G}3\,
    \rho+
    \frac1{2r_c^2}\left\{\,1-
    \sqrt{\,1-4r_c^2
    \left(\frac{\textstyle\Lambda_5}{\textstyle 6}
    +\frac{\textstyle\cal C}{\textstyle a^4}-
    \frac{\textstyle 8\pi G}{\textstyle 3}\,
    \rho\right)}\,\right\}                     \label{modFriedmannDGP}
    \end{eqnarray}
(note the negative sign of $\rho$ under the square root and the
extra first term on the right hand side) and in the limit of small
$\rho$, ${\cal C}/a^4$ and $\Lambda_5$ yields the behavior very
different from the GR limit of the anomaly driven model (\ref{GR}),
    \begin{eqnarray}
    \frac{\dot a^2}{a^2}+\frac{1}{a^2}\simeq
    \frac{\textstyle\Lambda_5}{\textstyle 6}
    +\frac{\textstyle\cal C}{\textstyle a^4}
    +r_c^2
    \left(\frac{\textstyle\Lambda_5}{\textstyle 6}
    +\frac{\textstyle\cal C}{\textstyle a^4}-
    \frac{\textstyle 8\pi G}{\textstyle 3}\,
    \rho\right)^2.
    \end{eqnarray}
For vanishing $\Lambda_5$ and ${\cal C}/a^4$ this behavior
corresponds to the 5D dynamical phase \cite{DGPDeffayet,DDG} which
is realized in the DGP model for a very small matter energy density
on the brane $\rho\ll 3/32\pi G r_c^2\sim m_P^2/r_c^2$.

Of course, in this range the DGP braneworld reduces to a vacuum
brane, but one can also imagine that the 5D cosmological constant
decays into matter constituents similar to (\ref{split}) and thus
simulates the effect of $\rho$ in Eq.(\ref{modFriedmann}). This can
perhaps provide us with a closer correspondence between the anomaly
driven cosmology and the non-vacuum DGP case. But here we would
prefer to postpone discussions of such scenarios to future analyses
and, instead, focus on the generalized {\em single-branch} DGP model
to show that it also admits the cosmological acceleration epoch
followed by the big boost singularity.

Indeed, for positive $\Lambda_5$ satisfying a very weak bound
    \begin{eqnarray}
    \Lambda_5>\frac3{2r_c^2} \label{bound}
    \end{eqnarray}
Eq.(\ref{modFriedmannDGP}) has a solution for which,  during the
cosmological expansion with $\rho\to 0$, the argument of the square
root vanishes and the acceleration tends to infinity (prime again
denotes the derivative with respect to $a$)
    \begin{eqnarray}
    \frac{\ddot a}a\simeq
    \frac{\textstyle\left[\,r_c^2\Big(
    \frac{\textstyle\Lambda_5}{\textstyle 6}
    +\frac{\textstyle\cal C}{\textstyle a^4}
    -\frac{\textstyle 8\pi G}{\textstyle 3}\,
    \rho
    \Big)\,\right]'}
    {\textstyle r_c^2\,\sqrt{\,1-4r_c^2
    \Big(
    \frac{\textstyle\Lambda_5}{\textstyle 6}
    +\frac{\textstyle\cal C}{\textstyle a^4}
    -\frac{\textstyle 8\pi G}{\textstyle 3}\,\rho
    \Big)}}\to\pm\infty.
    \end{eqnarray}
This is the big boost singularity labeled similarly to
(\ref{boostHubble}) by $\infty$ and having a finite Hubble factor
    \begin{eqnarray}
    &&\left(\frac{\textstyle\Lambda_5}{\textstyle 6}
    +\frac{\textstyle\cal C}{\textstyle a^4}-
    \frac{\textstyle 8\pi G}{\textstyle 3}\,
    \rho\right)_\infty
    =\frac1{4r_c^2},\\
    &&\left(\frac{\dot a^2}{a^2}
    +\frac{1}{a^2}\right)_\infty
    =\frac{\Lambda_5}6+\frac1{4r_c^2}.
    \end{eqnarray}

For the effective $a$-dependence of $r_c^2$ and $G\rho$ analogous to
(\ref{behavior}), $r_c^2(a)\sim a^n$ and $G\rho(a)\sim 1/a^3$, the
acceleration becomes positive at least for $n\geq 0$,
    \begin{eqnarray}
    \frac{\ddot a}a\simeq
    \frac{\textstyle
    n+ 32\pi G\,r_c^2\rho}
    {\textstyle 4r_c^2\,\sqrt{\,1+4r_c^2
    \Big(\frac{\textstyle 8\pi G}{\textstyle 3}\,
    \rho-\frac{\textstyle\Lambda_5}{\textstyle 6}
    -\frac{\textstyle\cal C}{\textstyle a^4}\Big)}}
    \to+\infty.
    \end{eqnarray}

Thus, the {\em single-branch} DGP cosmology can also lead to a big
boost version of acceleration. For that to happen, one does not
actually need a growing $r_c$  (which can be achieved at the price
of having a time dependent $G_5$ --- itself some kind of a modulus,
in a string inspired picture). The DGP crossover scale $r_c$ can be
constant, $n=0$, and the big boost singularity will still occur
provided the lower bound (\ref{bound}) is satisfied\footnote{Or,
more precisely, its small modification due to the dark radiation
contribution ${\cal C}/a^4$ which is very small at late stages of
expansion.}. When $\Lambda_5$ violates this bound, the acceleration
stage is eternal with the asymptotic value of the Hubble factor
squared $\dot
a^2/a^2=\big(1-\sqrt{1-2r_c^2\Lambda_5/3}\big)/2r_c^2$.

\section{Conclusions}
To summarize, we have obtained the modified Friedmann equation in
the anomaly driven cosmology with the microcanonical density matrix
initial conditions suggested in \cite{slih,why}. This equation
exhibits a gravitational screening of the quantum Casimir energy of
conformal fields --- this part of the total energy density does not
weigh, being degravitated due to the contribution of the conformal
anomaly. Also, in the low-density limit this equation does not only
recover the general relativistic behavior, but also establishes a
good correspondence with the dynamics of the Randall-Sundrum
cosmology via the AdS/CFT duality. Moreover, for very large and
rapidly growing value of the Gauss-Bonnet coefficient $\beta$ in the
conformal anomaly this equation features a regime of cosmological
acceleration followed by a big boost singularity. At this
singularity the scale factor acceleration grows in finite proper
time up to infinity with a finite limiting value of the Hubble
factor, when the Universe again enters a quantum phase demanding for
its description an UV completion of the low-energy semiclassical
theory.

A natural mechanism of growing $\beta$ can be based on the idea of
adiabatically evolving scale associated with extra dimensions
\cite{why} and realized within the picture of AdS/CFT duality,
according to which the conformal field theory is induced on the 4D
brane from the 5D non-conformal theory in the bulk. As is well
known, this duality underlies the justification of the 4D general
relativistic limit in the Randall-Sundrum model \cite{Gubser,HHR}.
Here we observed an extended status of this duality from the
cosmological perspective
--- the generalized Randall Sundrum model with the
Sschwarzschild-AdS bulk was shown to be equivalent to the anomaly
driven cosmology for small energy density. In particular, the
radiation contents of the latter was shown to be equivalent to the
dark radiation term ${\cal C}/a^4$ pertinent to the Randall-Sundrum
braneworld with a bulk black hole of mass ${\cal C}$ (well-known
duality of the bulk black hole and the thermal CFT on the brane
\cite{bulkBH}).

It is interesting that the initial conditions of anomaly driven
model establish the relation between the amount of radiation ${\cal
C}$ and the product of renormalized cosmological and gravitational
constants $G\Lambda\sim\Lambda/m_P^2$ --- the corollary of the
closed system of equations (\ref{efeq})-(\ref{bootstrap})
\cite{slih}. Such a relation is not known in the standard $SU(N)$
AdS/CFT version of the Randall-Sundrum scenario valid for the
effective $\Lambda=0$ and large $N^2\sim N_{\rm cdf}\sim\beta\gg 1$.
But this is consistent with the fact that the solution of the
bootstrap equations (\ref{efeq})-(\ref{bootstrap}) has a scaling
behavior in $N_{\rm cdf}\sim N^2$, $\Lambda\to\Lambda/N_{\rm cdf}$,
${\cal C}\to N_{\rm cdf}{\cal C}$ \cite{slih}, which simply implies
the limit of $N\to\infty$ in this scenario. This limit justifies the
semiclassical approximation applicable in accordance with
(\ref{landscape}) in the range of curvature much below the Planck
scale.

Another intriguing observation concerns establishing the {\em exact}
correspondence between the anomaly driven cosmology and the vacuum
DGP model generalized to the case of a nonvanishing $\Lambda_5$. In
this case a large $\beta$ is responsible for the large crossover
scale $r_c$, (\ref{DGPscale}). For positive $\Lambda_5$ satisfying
the lower bound (\ref{bound}) this model also features a big boost
scenario even for stabilized $\beta$. Below this bound (but still
for positive $\Lambda_5>0$, because a negative $\Lambda_5$ would
imply the point of maximal expansion from which the Universe starts
recollapsing) the cosmological evolution eventually enters eternal
acceleration scenario. However, the DGP model with a matter on the
brane can hardly be equivalent to the 4D anomaly driven cosmology,
unless one has some mechanism of decaying $\Lambda_5$ and simulating
matter density on the brane.

Unfortunately, the AdS/CFT correspondence with adiabatically
evolving scale of extra dimension cannot incorporate the
phenomenology of the observable dark energy, because the local
gravitational physics of this model becomes very different from the
4D general relativity. Thus, macabre perspective of being ripped by
infinite tidal forces at Big Boost singularity seems being
postponed. AdS/CFT-correspondence --- a perfect nonperturbative tool
in mathematical physics --- still remains a thing in itself from the
perspective of applied astroparticle cosmology.

In general, the idea of a very large central charge of CFT algebra,
underlying the solution of the hierarchy problem in the dark energy
phenomenon and particle phenomenology, seems hovering in current
literature \cite{bigcentalcharge,Dvalispecies}. Our idea of a big
growing $\beta$ belongs to the same scope, but its realization seems
escaping phenomenological framework. Probably some other
modification of this idea can be more productive. In particular,
another qualitative mechanism of running $\beta$ could be based on
the field-theoretic implementation of winding modes. These modes do
not seem to play essential role in the AdS/CFT picture with a big
scale of extra dimensions $L$, because they are heavy in the big $L$
limit. On the contrary, this mechanism is expected to work in the
opposite case of contracting extra dimensions, for which the
restrictions imposed by local gravitational physics do not seem to
apply (as long as for $L\to 0$ the short-distance correction go
deeper and deeper into UV domain). We hope to consider the mechanism
of winding modes in accelerating cosmology elsewhere.

\section*{Acknowledgements}
A.B. is grateful for hospitality of the Laboratory APC, CNRS,
University Paris 7, Paris, where a major part of this work has been
done. His work was also supported by the Russian Foundation for
Basic Research under the grant No 05-01-00996 and the grant
LSS-4401.2006.2. A.Yu.K. was supported by the RFBR grant 05-02-17450
and the grant LSS-1157.2006.2. A.B. and C.D. wish to thank
G.Gababadze and R.Woodard for discussions.

\end{document}